\documentclass{aa}  

\usepackage{graphicx}
\usepackage{txfonts}
\usepackage{gensymb}
\begin{document}

   \title{MUSE observations of comet 67P/Churyumov-Gerasimenko}

   \subtitle{A reference for future comet observations with MUSE}

   \author{
   C. Opitom\thanks{E-mail: copi@roe.ac.uk}
         \inst{1}\inst{2}
          \and
          A. Guilbert-Lepoutre\inst{3}
          \and
          S. Besse \inst{4}
          \and         
          B. Yang\inst{1}
           \and
          C. Snodgrass\inst{2}
          }
        \institute{ESO (European Southern Observatory) - Alonso de Cordova 3107, Vitacura, Santiago Chile
             \and
        Institute for Astronomy, University of Edinburgh, Royal Observatory, Edinburgh EH9 3HJ, UK
            \and
        Univ Lyon, Univ Lyon 1, ENSL, CNRS, LGL-TPE, F-69622, Villeurbanne, France
             \and
        Operations Department, European Space Astronomy Centre/ESA, P.O. Box 78, 28691 Villanueva de la Canada, Madrid, Spain}

   \date{}

% \abstract{}{}{}{}{} 
% 5 {} token are mandatory
 
  \abstract
   {}
   {Observations of comet 67P/Churyumov-Gerasimenko were performed with MUSE at large heliocentric distances post-perihelion, between March 3 and 7, 2016. Those observations were part of a simultaneous ground-based campaign aimed at providing large-scale information about comet 67P that complement the ESA/Rosetta mission.}
   {We obtained a total of 38 datacubes over 5 nights. We take advantage of the integral field unit (IFU) nature of the instrument to study simultaneously the spectrum of 67P's dust and its spatial distribution in the coma. We also look for evidence of gas emission in the coma.}
   {We produce a high quality spectrum of the dust coma over the optical range that could be used as a reference for future comet observations with the instrument. The slope of the dust reflectivity is of 10\%$/100$ nm over the 480-900 nm interval, with a shallower slope towards redder wavelengths. We use the $\mathrm{Af\rho}$ to quantify the dust production and measure values of 65$\pm$4 cm, 75$\pm$4 cm, and 82$\pm$4 cm in the V, R, and I bands respectively. We detect several jets in the coma, as well as the dust trail. Finally, using a novel method combining spectral and spatial information, we detect the forbidden oxygen emission line at 630 nm. Using this line we derive a water production rate of $1.5\pm0.6 \times 10^{26} \mathrm{molec./s}$, assuming all oxygen atoms come from the photo-dissociation of water.}
   {}

   \titlerunning{67P MUSE}
   \maketitle
%
%-------------------------------------------------------------------

\section{Introduction}
The ESA/Rosetta mission was the first mission  to approach and follow a comet for a large part of its orbit, as well as to land on its nucleus. The Rosetta orbiter and the Philae lander were equipped with a series of instruments to study in-situ the gas and dust coma surrounding comet 67P/Churyumov-Gerasimenko (hereafter 67P) and the properties of its nucleus. The mission provided unprecedented insights into the structure, composition, and activity evolution of 67P, and comets in general \citep{Taylor2017}. 

In order to complement the mission, an ambitious ground-based campaign was put together, with dozens of telescopes across the globe and in space performing observations of the comet at the same time as the Rosetta mission \citep{Snodgrass2017}. Those observation probed the large scale coma of 67P and provided context to the measurements performed by the Rosetta mission. More importantly, they provide a link with ground-based observations of a large number of comets that have been performed over the last century. As part of this campaign, observations were performed with the Multi Unit Spectroscopic Explorer (MUSE) instrument at the Very Large Telescopes (VLT). The observations were performed in March 2016, after the comet's perihelion passage, when the comet was moving away from the Sun. In this paper, we present the results of the MUSE observations of comet 67P.  

%--------------------------------------------------------------------
\section{Observations and data reduction}

MUSE is an integral field unit spectrograph mounted on the UT4 telescope of the VLT, in Chile \citep{Bacon2010}, nominally covering the 480-930 nm range. In Wide Field mode, which was used for our observations, MUSE has a field of view (FoV) of 1\arcmin$\times$1\arcmin$ $ covered without gaps. The instrument has a platescale of 0.2\arcsec /pix and a spectral resolving power between 1770 (at 480 nm) and 3590 (at 930 nm).
Observations of comet 67P were performed between 2016 March 3 and March 7. In total, 38 datacubes were obtained over 5 nights. For 4 of those cubes, the comet was not visible, or very badly centred, so that we discarded the data. The sky during those 5 nights was either clear or photometric, but the seeing was variable, between 0.6 and 2.3\arcsec. For all observations, we used an exposure time of 600s, and the position angle of the instrument was set to $0\degree$ (we did not apply any rotation between the exposures). At the time of the observations, even though the comet was still active, and thus extended, it did not fill the entire MUSE FoV. Because of that, no dedicated sky observations were performed. The observing circumstances are presented in Table \ref{TableObs}.

\begin{table*}
\caption{\label{TableObs} Observing circumstances of the 67P MUSE campaign}
\centering
\begin{tabular}{lcccccc}
\hline\hline
Date (UT) & N & Exposure time (s) & r (au) & $\Delta$ (au) & Airmass & Phase Angle (\degree)\\
\hline
2016-03-03 03h27-06h08 & 12 & 600 & 2.49 & 1.52 & 1.2-1.6 & 6 \\
2016-03-04 05h34 & 1 & 600 & 2.50 & 1.53 & 1.2 & 6 \\
2016-03-05 04h50-06h29 & 8 & 600 & 2.51 & 1.53 & 1.2-1.3 & 5\\
2016-03-06 03h25-04h00 & 4 & 600 & 2.52 & 1.54 & 1.4-1.5 & 5 \\
2016-03-07 04h39-07h38 & 13 & 600 & 2.52 & 1.54 & 1.3-1.6 & 4\\
\hline
\end{tabular}
%\tablefoot{}
\end{table*}

The data reduction was performed using the ESO pipeline \citep{Weilbacher2016}, with the sky estimated from regions near the edge of the cubes that are free from the comet contamination. In addition to the cube reconstruction, bias subtraction, flatfield correction, wavelength calibration, and sky subtraction, the ESO MUSE pipeline also corrects for the telluric absorption and flux-calibrates the science data, using a standard star observed the same night as the science observations. 

Even though the sky was estimated directly on the science cubes and subtracted by the pipeline, while examining the reduced cubes we noticed relatively strong sky residuals. In order to reduce those, we then used the ZAP (Zurich Atmosphere Purge) software \citep{Soto2016}. ZAP is a Principal Component Analysis-based software designed to perform sky subtraction on IFU data. The use of ZAP allows us to remove most of the sky residuals left after the pipeline reduction. As will be discussed later, we also performed a full data reduction without applying any sky subtraction, in order to search for forbidden oxygen emission lines in the coma of 67P.

As mentioned before, the ESO pipeline performs a correction of the telluric absorption using a standard star observed on the same night as the science data. However, the standard stars are primarily  used for flux calibration and are not optimal for telluric correction. Also, they are not always observed just after the science observations nor at the same airmass. Because of that, in the reduced cubes, there are residuals of the strong O$_2$ telluric band around 760 nm if the telluric correction is performed with the pipeline. For all the analysis focused on the 2D structure of the dust coma, and the detection of forbidden oxygen lines, this does not impact the quality of our measurements, as those residuals are relatively constant over the field and we extracted the cubes over wavelength ranges that mostly avoid that region of the spectrum. However, in Section \ref{Ref_Spectrum}, we focus on the spectrum of the dust in the coma of 67P and residual from the telluric correction could impact the quality of the spectrum presented. For the spectra presented in that section, we thus used the Molecfit software \citep{Smette2015,Kausch2015} to perform a better telluric correction. Molecfit has proven to be very useful to provide an accurate telluric correction even when dedicated observations of a standard star were not available. We applied Molecfit directly on the extracted 1D spectra. The fit of the telluric features was performed for each extracted spectrum individually.

\section{Analysis}

\subsection{A reference dust spectrum}
\label{Ref_Spectrum}

   \begin{figure*}[h!]
   \centering
   \includegraphics{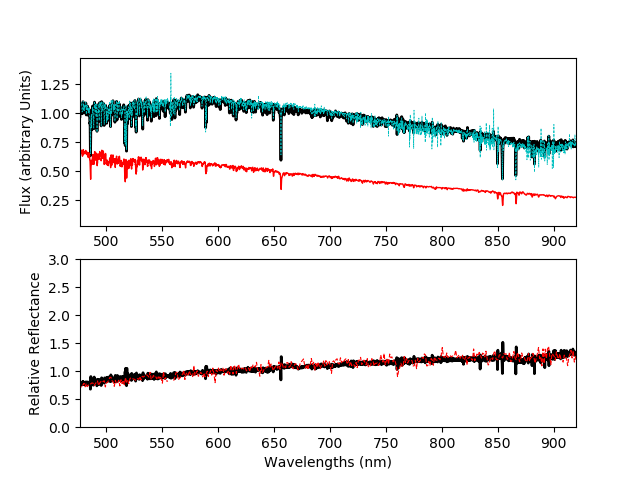}
   \caption{Top: Spectrum of 67P dust coma (black), together with the solar spectrum used to compute the reflectance spectrum (red). The blue dotted line is the spectrum extracted over a 10\arcsec$ $ aperture.  The solar spectrum has been shifted arbitrarily over the y-axis for better visibility. Bottom: Relative reflectance spectrum of 67P (black), compared to the one measured with X-Shooter in November 2014 \citep[red;][]{Snodgrass2016}.}
    \label{Dust_Spect}
    \end{figure*}

We first study the spectrum of comet 67P over the optical range. To do so, we extract all the data cubes over a 5 pixels radius (1\arcsec) aperture around the comet optocenter. We chose such a small aperture to focus on the part of the coma where the signal is the strongest, to avoid having strong sky residuals in the presented spectrum. After the extraction, the spectra are corrected for the telluric features using the Molecfit software, as explained above. All 34 spectra are then median-combined to produce a high quality spectrum of the dust in the coma of 67P. This spectrum is shown in the upper part of Fig. \ref{Dust_Spect}. We do not see any emission band in the spectrum, because of the large heliocentric distance and low activity level of the comet at the time of the observations. In the same part of the Figure, we show a reference solar spectrum obtained using the SOLar SPECtrometer (SOLSPEC) instrument of the SOLAR payload on board the International Space Station (ISS) \citep{Meftah2018}, re-sampled to match the sampling of the 67P spectrum. Given that the aperture we chose is very small, the spectrum shown in Fig. \ref{Dust_Spect} probably contains a non-negligible nucleus contribution. For comparison purposes, we also extract the dust spectrum over a 10\arcsec$ $ aperture, in which the dust coma dominates (blue dashed curve in the top part of Fig. \ref{Dust_Spect}). The two spectra match well, except for a small difference around 880 nm. This region is severely affected by telluric absorption and has strong sky emissions, that might explain this difference. In general, we can thus say that the spectrum extracted over the 1\arcsec$ $ aperture is representative of the coma dust and we will discuss this spectrum only in the following text. 

In the bottom part of Fig. \ref{Dust_Spect}, we divide the comet spectrum by the Solar reference spectrum to compute the relative reflectance of the dust in the coma of 67P. The reflectance spectrum is normalised at 600 nm. Overlaid to the reflectance spectrum measured with MUSE, is the reflectance measured using the X-Shooter spectrograph in November 2014 over the same wavelength range \citep{Snodgrass2016}. The reflectance spectra measured in November 2014 and March 2016 are consistent with each other, indicating that the dust reflectance as measured from the ground is similar at large heliocentric distance pre- and post-perihelion. We do not see any sign of absorption bands in the optical spectrum of 67P. We measure a spectral slope for the reflectance, or dust reddening, of 10\%$/100$ nm in the 480-900 nm interval. We also notice that the slope becomes shallower at longer wavelengths. We measure a slope of 13 \%$/100$ nm in the 500-700 nm interval but only of 5\%$/100$ nm in the 700-900 nm interval. This is fully consistent with what is reported from X-Shooter observations performed in 2014, from which values between 10\%$/100$ nm and 20\%$/100$ nm were reported in the 550 to 1000 nm interval, the shallower values corresponding to the red end of the wavelength range. 

Our measurements are also consistent with in-situ measurements of the dust from the ESA/Rosetta mission. \cite{Bertini2017} report a reddening in the interval 376-744 nm ranging between 11 and 14\%$/100$ nm from measurements with the OSIRIS cameras. Similarly, \cite{LaForgia2019} report slopes measured in the inner coma ranging from 12 to 16\%$/100$ nm between 480 and 649 nm. Those slopes are similar to those measured for the nucleus of 67P, both in-situ and from ground-based observations. \cite{Fornassier2015} report average slopes of 11$-$16\%$/100$ nm over 250$-$1000 nm with shallower slopes toward longer wavelengths, from observation of 67P's nucleus with the Rosetta/OSIRIS camera. From ground-based observations, \cite{Tubiana2011} measure a slope for the nucleus of $12\pm1$\%$/100$ nm over 430$-$850 nm, slightly shallower in the 500$-$850 nm range. Finally, the dust reddening measured in the coma of 67P is consistent with what is measured usually in the coma of active comets, typically between 0 and 20\%$/100$ nm, and with a shallower slope towards the near-IR \citep{Solontoi2012,Jewitt1986}.

\subsection{Dust coma morphology and activity}
\label{Morphology}

\begin{figure*}[h!]
\centering
\includegraphics[width=5.2cm,height=5.2cm]{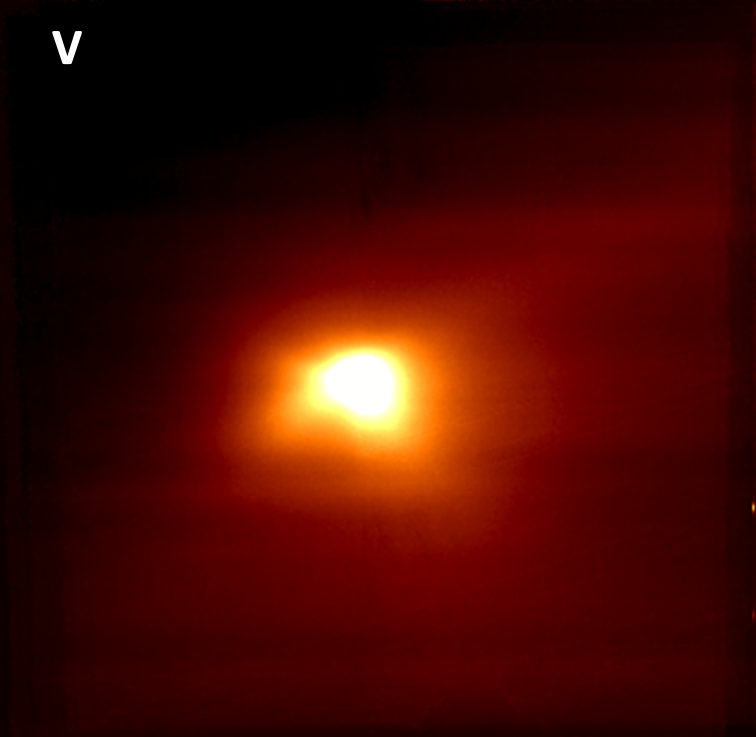}
\includegraphics[width=5.2cm,height=5.2cm]{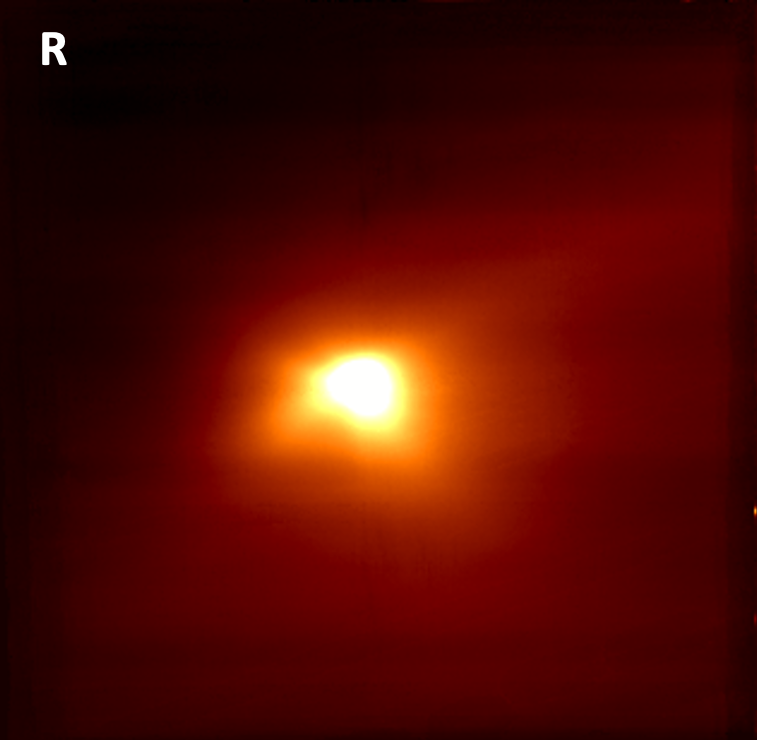}
\includegraphics[width=5.2cm,height=5.2cm]{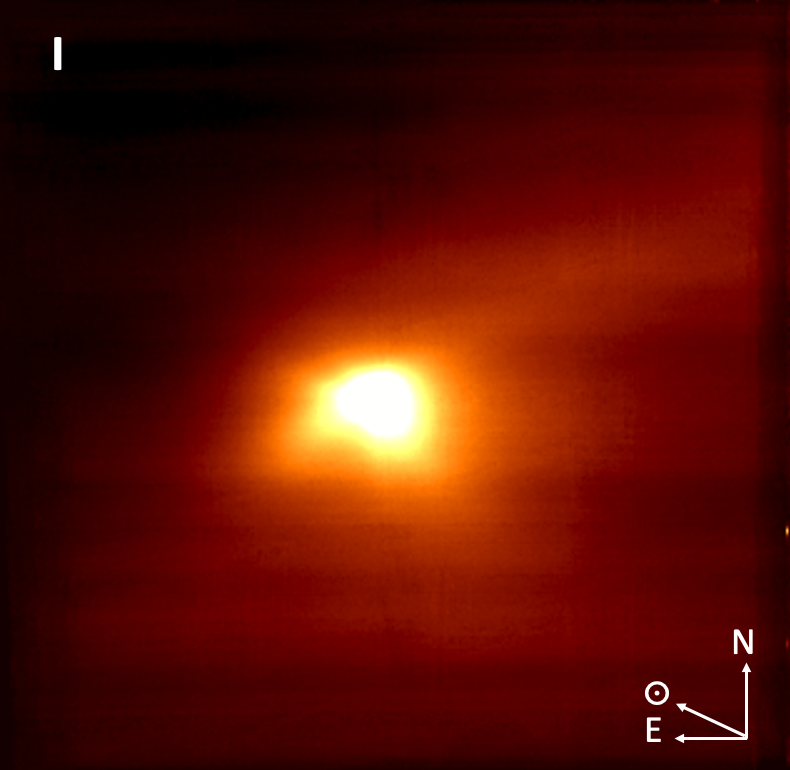}
\includegraphics[width=5.2cm,height=5.2cm]{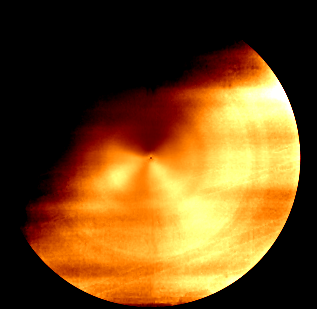}
\includegraphics[width=5.2cm,height=5.2cm]{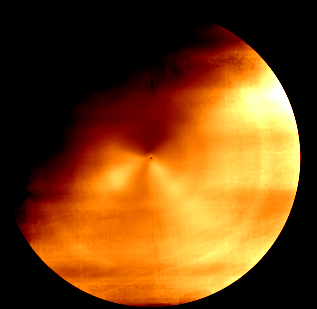}
\includegraphics[width=5.2cm,height=5.2cm]{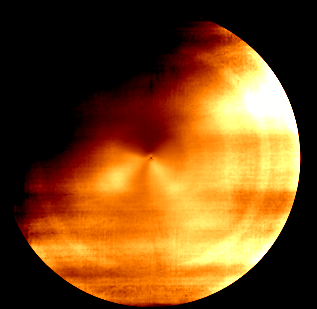}
\caption{Top row: combined maps of 67P in V, R, and I bands. The maps are centred on the comet and the FoV is 1\arcmin$\times$1\arcmin. Bottom row: V, R, and I maps enhanced by dividing by an azimuthal median profile \citep{Samarasinha2013}.}
\label{DustMaps}
\end{figure*}

For each cube in our dataset, we extract maps over the bandpasses of the V, R, and I Johnson-Cousins filters. All maps in the same bandpass are re-centred and then co-added. The resulting co-added maps are displayed in the top part of Fig. \ref{DustMaps}. The dust coma morphology is the same for all three bandpasses. It is asymmetrical, probably due to the presence of dust jets. To investigate further the presence of jets, we divide the dust maps by an azimuthal median profile \citep{Samarasinha2013}. Enhanced maps are displayed on the bottom part of Fig. \ref{DustMaps}. On those we can clearly see two jets, which are most likely the cause of the apparent asymmetry of the dust coma. The first jet is located close to the anti-sunward direction, towards the South-West. The second jet is located about 90\degree$ $ away, towards the South-East. Finally, we see a faint feature towards the sunward direction. Those jets are consistent with what was observed on previous passages and what is reported by \cite{Knight2017} and \cite{Snodgrass2017} from observations at the same epoch. They are also consistent with the modelling of the pole orientation and active region location done by \cite{Vincent2013} prior to the Rosetta mission. In addition to the two jets mentioned above, we see an enhancement towards the North-West. This corresponds to the dust trail, that was reported to be at least two degrees long at that epoch \citep{Snodgrass2017,Boehnhardt2016,Knight2017}. We do not see changes of the coma morphology over the 5 nights during which we have observations, nor over a single night.
%\cite{Knight2017} also attempt to link the active regions identified from the observations of large-scale jets in the coma to the specific morphology of the nucleus as seen from the Rosetta mission. 

To constrain the comet activity at that time, we compute the Af$\rho$ value, which is a proxy for dust production, as defined by \cite{A'Hearn1984}. We compute the $\mathrm{Af\rho}$ for the V, R, and I bands over a 2500 km physical aperture. We use a 2500 km aperture instead of the more commonly used 10000 km aperture because the comet signal drops significantly at 10000 km. We obtain values of 65$\pm$4 cm, 75$\pm$4 cm, and 82$\pm$4 cm in the V, R, and I bands respectively. Those values have not been corrected for the phase angle effect and are consistent with those reported by \cite{Boehnhardt2016} at the same epoch, and comparable (even though slightly lower) to those reported by \cite{Knight2017} once the phase angle effect is taken into account. Using the Af$\rho$ values in the different bands, we can also compute the dust reflectivity gradient defined as \citep{Jewitt1986}:

\[ \mathrm{S' = \frac{Af\rho_2- Af\rho_1}{Af\rho_2+ Af\rho_1} \frac{20000}{\lambda_2-\lambda_1}}\]

where $\mathrm{Af\rho_x}$ and the corresponding wavelength $\lambda_x$ are expressed in cm and nm, respectively. For the I-V combination, this gives a reflectivity gradient of 7$\pm$1 \%$/100$ nm, while it is of 11$\pm$2 \%$/100$ nm for the I-R and 7$\pm$1 \%$/100$ nm for the R-V combination. This confirms the trend outlined in section \ref{Ref_Spectrum} of higher reflectivity gradient at lower wavelengths. To check for trends with the aperture size, we computed the Af$\rho$ in apertures of 5000 km, 7500 km, and 10000 km. Within the error bars, we do not see significant changes in the $\mathrm{Af\rho}$ or reflectivity gradient values with the aperture size in our data, as expected for a steady-state coma.

\subsection{Gas detection}
\label{Gas}
Among the features usually observed in the optical spectrum of comets, the oxygen forbidden line at 630 nm is one of the brightest. There are three forbidden oxygen lines that can be detected in the coma of comets at optical wavelengths: the green line at 557.7 nm, and the red doublet at 630 and 636.4 nm. Those lines are emitted by the decay of oxygen atoms in a metastable ($^1$D) or ($^1$S) state. Exited atomic oxygen is mainly produced through the photo-dissociation of $\mathrm{H_2O}$, $\mathrm{CO}$, $\mathrm{CO_2}$, or even $\mathrm{O_2}$ (see, e.g. \cite{Cessateur2016}). Observing forbidden oxygen lines then represents an opportunity to constrain the production of those potential parent species from optical observations. However, the cometary forbidden oxygen lines are often blended with the equivalent atmospheric lines, unless observed at high spectral resolution.

The spectral resolution of MUSE does not allow us to resolve the telluric and cometary lines for the geocentric velocity of comet 67P at the time of our observations. In theory, the sky subtraction should subtract the atmospheric contribution, allowing us to recover the cometary signal. However, the noise introduced by the subtraction of the very strong atmospheric features prevents us from detecting any cometary forbidden oxygen emission lines. At the time of our observations the comet was at 2.5 au post-perihelion and was only weakly active, as confirmed by the fact that we do not detect any emission lines in the spectrum presented in Fig. \ref{Dust_Spect}. If the oxygen lines are present, they are thus very faint and masked by the noise introduced by the sky subtraction. 

\begin{figure*}[h!]

\includegraphics[width=18cm]{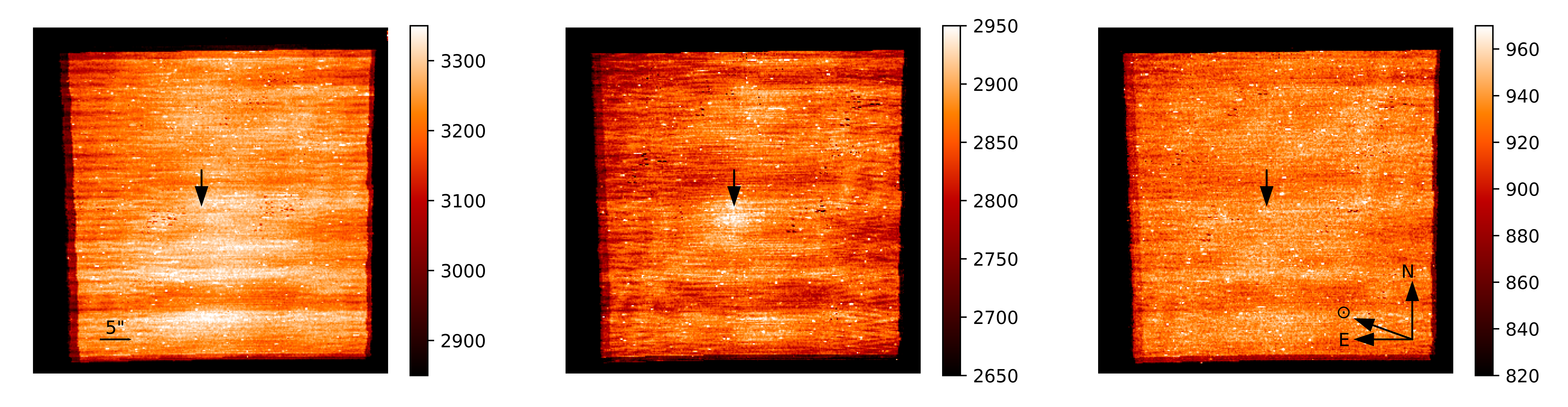}
\caption{Combined maps of the sky+comet flux at the wavelengths of the 3 forbidden oxygen lines at 557.7 nm (left), 630 nm (middle), and 636.4 nm (right). The FoV is 52\arcsec$\times$52\arcsec and is oriented North up and East left and the black arrow in the centre points to the position of the comet optocenter. The unit of the color bars is $10^{-20}$ erg/s/cm$^2$.}
\label{OMAPS}
\end{figure*}

We therefore attempt to detect the forbidden oxygen lines using another method. We reduce the full dataset without performing any sky subtraction. For each spaxel of each cube, we then subtract the continuum sky contribution and the dust contribution underlying the oxygen forbidden lines by defining continuum region on both sides of the red doublet and the green oxygen lines, fitting a line through those regions and subtracting it. Finally, we extract the datacubes over a very narrow wavelength range centred on the wavelength of the three oxygen lines. In low/medium resolution spectra, forbidden oxygen lines can be contaminated by $\mathrm{NH_2}$ or $\mathrm{C_2}$ lines, but this is highly unlikely in the case of those observations since the comet is weakly active and no emission lines are detected in Fig. \ref{Dust_Spect}. The result is maps of the flux contained in the forbidden lines at 557.7, 630, and 636.4 nm over the whole MUSE FoV. Since the atmospheric and cometary lines cannot be resolved by MUSE, the maps contain the sum of the atmospheric and potential cometary contribution. All maps are re-centred so that the optocenter of the comet is placed at the same position. We average all 34 maps for each line, performing a 3-$\sigma$ clipping. The result in shown in Fig. \ref{OMAPS}

In those maps, we expect the atmospheric contribution to be relatively uniform over the 1\arcmin$\times$1\arcmin$ $ FoV of MUSE. The comet contribution, on the other side, would be concentrated around the optocenter of the comet, with a limited spatial extension since the states responsible for the emission lines are metastable states. In Fig \ref{OMAPS}, we can see that for the 557.7 and 636.4 nm lines, the maps are mostly uniform over the whole FoV. Small variations are observed but they most likely come from non-optimal corrections of the detector-to-detector effects. We thus conclude that the signal in those maps comes from the atmospheric oxygen lines. In the 630 nm map, however, in addition to inhomogeneities similar to those seen in the other two maps, we have a clear over-density located around the position of the comet optocenter. This indicates that, in addition to the atmospheric line, we detect signal from the cometary forbidden oxygen line at 630 nm. It is not surprising that we only detect the 630 nm line, since it is the strongest of the three oxygen lines. The signal we measure in the comet aperture for the 630 nm map is barely 3-$\sigma$ above the sky background variation (measured from four adjacent sky apertures, see later). Since the ratio between the 630 nm and 636.4 nm maps is expected to be 3 (as it is the case for the background in the two maps, see Fig. \ref{OMAPS}), it is then consistent with the fact that we do not detect significant cometary signal in the 636.4 nm map. We note that the [OI] cometary signal is enhanced towards the South, similarly to what is observed for the dust. However, we do not believe the signal is due to residuals from the dust continuum. Indeed, the technique used for the dust subtraction is the same of all 3 lines and only the 630 nm map (with the strongest oxygen feature) shows an over-density around the optocenter. Control maps built by using a wavelength range adjacent to the 630 nm line do not show a similar signal.

The cometary signal we detect in the 630 nm map is faint, but we use this signal to estimate the water production rate of the comet at that time. The 630 nm line comes from the decay of atomic oxygen in ($^1$D) state. As mentioned before, oxygen in that state is mainly produced by the photo-dissociation of $\mathrm{H_2O}$, $\mathrm{CO}$, $\mathrm{CO_2}$, or $\mathrm{O_2}$. For comets at 1 au from the Sun, $\mathrm{H_2O}$ photodissociation is the dominant source for producing metastable oxygen, but at distance above 2.5 au other molecules such as CO and $\mathrm{CO_2}$ start to contribute \citep{Decock2013,McKay2015}. Long-term measurements of the production rate of all 4 species of interest derived from  the ROSINA instrument onboard Rosetta are presented by \cite{Laeuter2020} and \cite{Combi2020}. At the time of our observations, the CO and $\mathrm{O_2}$ production rates are more than a factor 10 and 100 lower than that of $\mathrm{H_2O}$, respectively. We can thus assume that those species will contribute little to the production of metastable oxygen. The $\mathrm{CO_2}$ production rate is only about a factor 2 to 4 lower than the $\mathrm{H_2O}$ production rate. The emission rate for O($^1$D) production from $\mathrm{CO_2}$ is 1.5 times higher than that of $\mathrm{H_2O}$ \citep{Bhardwaj2012}. Taking this into account, a significant part (up to half) of O($^1$D) atoms could be produced by the photo-dissociation of $\mathrm{CO_2}$. For the measurements presented below, we will assume that water is the main source for the production of metastable oxygen, so that the water production rate we derive might be overestimated.

In order to derive water production rates, we follow the procedure described in \cite{Schultz1992,Morgenthaler2001}. We consider a photo-chemical model including the following three reactions,
\begin{equation}
\mathrm{H_2O + h\nu \longrightarrow H_2 + O(^1D)}
\label{equ1}
\end{equation}
\begin{equation}
\mathrm{H_2O + h\nu \longrightarrow H + OH}
\label{equ2}
\end{equation}
\begin{equation}
\mathrm{OH + h\nu \longrightarrow H + O(^1D)}
\label{equ3}
\end{equation}
the water production rate is given by:

\[ \mathrm{Q(H_2O) = \frac{Q([OI])}{BR1+(BR2)(BR3)} }\]
where BR1 and BR2 are the branching ratios for reactions \ref{equ1} and \ref{equ2} \citep{Huebner1992} (equal to 0.05 and 0.855 for the quiet Sun, respectively) and BR3 is the branching for reaction \ref{equ3} from \cite{Morgenthaler2001} (equal to 0.094). The production rate of oxygen atoms in ($^1$D) state, Q([OI]), can be obtained from:
\[ \mathrm{Q([OI]) = \frac{4}{3}4\pi\Delta^2I_{630}AC }\]
where $\mathrm{\Delta}$ is the comet geocentric distance (in cm), $\mathrm{I_{630}}$ is the intensity of the [OI] emission (in $\mathrm{photons/s/cm^2}$), and AC is the aperture correction factor correcting for the [OI] emission not encompassed in the aperture.

We measure the comet flux in the 630 nm map using a 25 pixels (5\arcsec$ $) radius circular aperture centred on the comet optocentre. In order to measure and subtract the sky flux, we measure the flux in four 5\arcsec$ $ apertures located at four different positions with respect to the comet optocenter ((+50 pix,+50 pix), (+50 pix,-50 pix), (-50 pix,+50 pix), (-50 pix,-50 pix)), computed the average, and subtracted it from the flux measured in the comet aperture. Given the faintness of the cometary emission, the main uncertainty in the determination of the water production rate comes from the measurement of the sky. To estimate this uncertainty, we used the upper and lower values measured in the individual sky apertures. The resulting intensity of the comet 630 nm line is $1.8\pm1.0 \times 10^{-3}  \mathrm{photons/cm^2/s}$. The aperture we used encompasses all the visible emission from the comet, so that we set the aperture correction factor to 1. This results in a water production rate of $1.5\pm0.7 \times 10^{26} \mathrm{molec./s}$. This represents the actual water production rate only if all the oxygen atoms in ($^1$D) state are produced by the photo-dissociation of water. 

No other measurement of the comet water production rate was reported from ground-based observations at a similar epoch because the comet was too faint and weakly active. In-situ measurements made with the ROSINA DFMS instruments determined a water production rate around $2.3\times 10^{26}$ molec./s at the same heliocentric distance \citep{Hansen2016}. Similar values are reported by \cite{Combi2020,Laeuter2020} using the same instrument. \cite{Biver2019} report a water production rate of $8.5\pm2.5 \times 10^{25} \mathrm{molec./s}$ from measurements with the MIRO instrument. Given that $\mathrm{CO_2}$ could potentially contribute up to half of the 630 nm [OI] line brightness, the actual water production rate we measure could be as low as $0.7 \times 10^{26} \mathrm{molec./s}$, which is in very good agreement with the \cite{Biver2019} measurement. Given the difference in technique, scale of the observations and models used, our measurement is close to those reported using Rosetta instruments in particular the MIRO instrument.

\section{Discussion and Conclusions}

\begin{figure*}[h!]
\centering
\includegraphics[width=17.2cm]{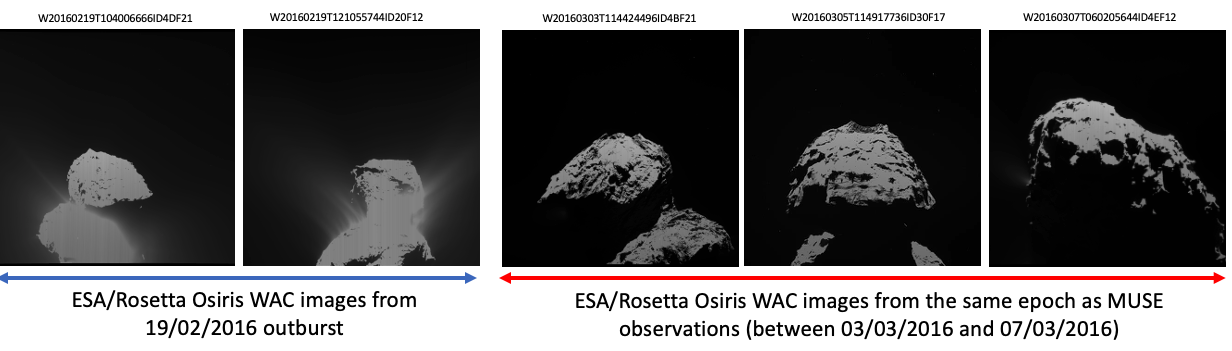}
\caption{Images of the nucleus of 67P taken by the OSIRIS instrument on-board Rosetta during the 19/02/2016 outburst and at the same epoch as the MUSE observations presented here. Credit: ESA/Rosetta/MPS for OSIRIS Team MPS/UPD/LAM/IAA/SSO/INTA/UPM/DASP/IDA.}
\label{Plot_Rosetta}
\end{figure*}

Comet 67P was observed with the MUSE IFU over 5 nights, between March 3 and 7, 2016 when the comet was at 2.5 au from the Sun and 1.5 au from the Earth. The spectrum of the dust coma of 67P presented here is of high quality. It has a good signal to noise ratio given the faintness and the distance of the comet at the time of our observations. Due to the use of the Molecfit software it is little affected by telluric features at near-IR wavelengths, which is not the case of most comet spectra in the 800-900 nm range. It matches very well other ground-based observations of 67P, as well as in-situ measurements from the Rosetta spacecraft. It is also representative of what is usually observed for active comets in general. Finally, it does not contain gas emission features above the noise level (some [OI] signal is present, merged with the sky line; see below). For all those reasons, this spectrum could be used in the future as a 'template' dust spectrum to help perform the subtraction of the dust-reflected continuum for observations of other comets with the MUSE IFU. This spectrum could replace the observation of solar analogs since so far very few good solar analogs have been observed with MUSE. The dust reddening can vary from one comet to another, so that this spectrum would need to be corrected from any slope difference between the target and 67P. Nonetheless, this dust-only spectrum of 67P represents a good tool to help for the analysis of future comet observations with MUSE.

The morphology of the coma, the dust activity, and the dust reflectivity gradient as measured with MUSE are consistent with other measurements performed with ground-based telescopes at the same epoch as well as with in-situ measurements from the Rosetta mission. 

About two weeks before our observations, on February 2 2016, an outburst was detected by several instruments on-board the Rosetta orbiter \citep{2016Grun}. A similar but less intense event was also detected in NAVCAM images on 1 March 2016. Since the outburst(s) happened shortly before our observations, we could have expected to detect some indications of such event, either in terms of activity level or coma morphology, in our observations. However, we do not detect any sign of the outburst in the MUSE data presented here. In Fig. \ref{Plot_Rosetta}, we show images obtained with the OSIRIS WAC on-board Rosetta. In the left part, gas jets are easily detected during the February 19 outburst. The right side shows the comet at the same epoch as our observations. No gas jets are visible. This is consistent with the comet being back to a quieter state and the fact that we do not detect the outburst in the MUSE observations. 

Through careful reprocessing and separating comet and sky signals spatially (rather than by resolving them spectroscopically), we detect the 630 nm forbidden oxygen line and derive a water production rate of $1.5\pm0.7 \times 10^{26} \mathrm{molec./s}$ in the coma of 67P, if all [OI] atoms in ($^1$D) state are produced by water. This value is consistent with the Rosetta measurements \citep{Hansen2016,Combi2020,Laeuter2020,Biver2019}. It is the only measurement of the water production rate of 67P at such large heliocentric distance from remote observations. Ground-based detection of water (or water products) in the coma of 67P only happened close to perihelion \citep{Snodgrass2017}, so that this point is precious to compare measurements and models from the Rosetta spacecraft to ground-based measurements over a larger portion of the comet orbit. In general, detecting water at such low production rate for comets at 2.5 au from the Sun is extremely difficult (see e.g. discussion on water detection in \cite{Snodgrass2017b}). 
MUSE had been shown in the past to have a huge potential to study species parentage in the coma of brighter comets \citep{Opitom2019}.
Our present work demonstrate the efficiency of MUSE to detect low level of water production around distant solar system bodies. This opens up future opportunities, at a time where an intensive search for evidence of water ice in the main asteroid belt and elsewhere in the Solar System is being performed.

\begin{acknowledgements}
 Based on observations made with ESO Telescopes at the La Silla Paranal Observatory under programme 096.C-0160(A). Datasets of the ESA/Rosetta OSIRIS instrument have been downloaded from the ESA Planetary Science Archive. The authors acknowledge the OSIRIS Principal Investigator H. Sierks (MPS, Goettingen, Germany) and the OSIRIS Team for providing images and related datasets and the ESA Rosetta Project for enabling the science of the mission.

- 
\end{acknowledgements}

\bibliographystyle{aa} % style aa.bst
\bibliography{Biblio} % your references Yourfile.bib

\begin{thebibliography}{34}
\expandafter\ifx\csname natexlab\endcsname\relax\def\natexlab#1{#1}\fi

\bibitem[{{A'Hearn} {et~al.}(1984){A'Hearn}, {Schleicher}, {Millis}, {Feldman},
  \& {Thompson}}]{A'Hearn1984}
{A'Hearn}, M.~F., {Schleicher}, D.~G., {Millis}, R.~L., {Feldman}, P.~D., \&
  {Thompson}, D.~T. 1984, \aj, 89, 579

\bibitem[{{Bacon} {et~al.}(2010){Bacon}, {Accardo}, {Adjali}, {Anwand},
  {Bauer}, {Biswas}, {Blaizot}, {Boudon}, {Brau-Nogue}, {Brinchmann},
  {Caillier}, {Capoani}, {Carollo}, {Contini}, {Couderc}, {Daguis{\'e}},
  {Deiries}, {Delabre}, {Dreizler}, {Dubois}, {Dupieux}, {Dupuy}, {Emsellem},
  {Fechner}, {Fleischmann}, {Fran{\c c}ois}, {Gallou}, {Gharsa}, {Glindemann},
  {Gojak}, {Guiderdoni}, {Hansali}, {Hahn}, {Jarno}, {Kelz}, {Koehler},
  {Kosmalski}, {Laurent}, {Le Floch}, {Lilly}, {Lizon}, {Loupias}, {Manescau},
  {Monstein}, {Nicklas}, {Olaya}, {Pares}, {Pasquini}, {P{\'e}contal-Rousset},
  {Pell{\'o}}, {Petit}, {Popow}, {Reiss}, {Remillieux}, {Renault}, {Roth},
  {Rupprecht}, {Serre}, {Schaye}, {Soucail}, {Steinmetz}, {Streicher}, {Stuik},
  {Valentin}, {Vernet}, {Weilbacher}, {Wisotzki}, \& {Yerle}}]{Bacon2010}
{Bacon}, R., {Accardo}, M., {Adjali}, L., {et~al.} 2010, in \procspie, Vol.
  7735, Ground-based and Airborne Instrumentation for Astronomy III, 773508

\bibitem[{{Bertini} {et~al.}(2017){Bertini}, {La Forgia}, {Tubiana},
  {G{\"u}ttler}, {Fulle}, {Moreno}, {Frattin}, {Kovacs}, {Pajola}, {Sierks},
  {Barbieri}, {Lamy}, {Rodrigo}, {Koschny}, {Rickman}, {Keller}, {Agarwal},
  {A'Hearn}, {Barucci}, {Bertaux}, {Bodewits}, {Cremonese}, {Da Deppo},
  {Davidsson}, {Debei}, {De Cecco}, {Drolshagen}, {Ferrari}, {Ferri},
  {Fornasier}, {Gicquel}, {Groussin}, {Gutierrez}, {Hasselmann}, {Hviid}, {Ip},
  {Jorda}, {Knollenberg}, {Kramm}, {K{\"u}hrt}, {K{\"u}ppers}, {Lara},
  {Lazzarin}, {Lin}, {Moreno}, {Lucchetti}, {Marzari}, {Massironi}, {Mottola},
  {Naletto}, {Oklay}, {Ott}, {Penasa}, {Thomas}, \& {Vincent}}]{Bertini2017}
{Bertini}, I., {La Forgia}, F., {Tubiana}, C., {et~al.} 2017, \mnras, 469, S404

\bibitem[{{Bhardwaj} \& {Raghuram}(2012)}]{Bhardwaj2012}
{Bhardwaj}, A. \& {Raghuram}, S. 2012, \apj, 748, 13

\bibitem[{{Biver} {et~al.}(2019){Biver}, {Bockel{\'e}e-Morvan}, {Hofstadter},
  {Lellouch}, {Choukroun}, {Gulkis}, {Crovisier}, {Schloerb}, {Rezac}, {von
  Allmen}, {Lee}, {Leyrat}, {Ip}, {Hartogh}, {Encrenaz}, {Beaudin}, \& {Miro
  Team}}]{Biver2019}
{Biver}, N., {Bockel{\'e}e-Morvan}, D., {Hofstadter}, M., {et~al.} 2019, \aap,
  630, A19

\bibitem[{{Boehnhardt} {et~al.}(2016){Boehnhardt}, {Riffeser}, {Kluge}, {Ries},
  {Schmidt}, \& {Hopp}}]{Boehnhardt2016}
{Boehnhardt}, H., {Riffeser}, A., {Kluge}, M., {et~al.} 2016, \mnras, 462, S376

\bibitem[{{Cessateur} {et~al.}(2016){Cessateur}, {de Keyser}, {Maggiolo},
  {Gibbons}, {Gronoff}, {Gunell}, {Dhooghe}, {Loreau}, {Vaeck}, {Altwegg},
  {Bieler}, {Briois}, {Calmonte}, {Combi}, {Fiethe}, {Fuselier}, {Gombosi},
  {H{\"a}ssig}, {Le Roy}, {Neefs}, {Rubin}, \& {S{\'e}mon}}]{Cessateur2016}
{Cessateur}, G., {de Keyser}, J., {Maggiolo}, R., {et~al.} 2016, Journal of
  Geophysical Research (Space Physics), 121, 804

\bibitem[{{Combi} {et~al.}(2020){Combi}, {Shou}, {Fougere}, {Tenishev},
  {Altwegg}, {Rubin}, {Bockel{\'e}e-Morvan}, {Capaccioni}, {Cheng}, {Fink},
  {Gombosi}, {Hansen}, {Huang}, {Marshall}, \& {Toth}}]{Combi2020}
{Combi}, M., {Shou}, Y., {Fougere}, N., {et~al.} 2020, \icarus, 335, 113421

\bibitem[{{Decock} {et~al.}(2013){Decock}, {Jehin}, {Hutsem{\'e}kers}, \&
  {Manfroid}}]{Decock2013}
{Decock}, A., {Jehin}, E., {Hutsem{\'e}kers}, D., \& {Manfroid}, J. 2013, \aap,
  555, A34

\bibitem[{{Fornasier} {et~al.}(2015){Fornasier}, {Hasselmann}, {Barucci},
  {Feller}, {Besse}, {Leyrat}, {Lara}, {Gutierrez}, {Oklay}, {Tubiana},
  {Scholten}, {Sierks}, {Barbieri}, {Lamy}, {Rodrigo}, {Koschny}, {Rickman},
  {Keller}, {Agarwal}, {A'Hearn}, {Bertaux}, {Bertini}, {Cremonese}, {Da
  Deppo}, {Davidsson}, {Debei}, {De Cecco}, {Fulle}, {Groussin}, {G{\"u}ttler},
  {Hviid}, {Ip}, {Jorda}, {Knollenberg}, {Kovacs}, {Kramm}, {K{\"u}hrt},
  {K{\"u}ppers}, {La Forgia}, {Lazzarin}, {Lopez Moreno}, {Marzari}, {Matz},
  {Michalik}, {Moreno}, {Mottola}, {Naletto}, {Pajola}, {Pommerol}, {Preusker},
  {Shi}, {Snodgrass}, {Thomas}, \& {Vincent}}]{Fornassier2015}
{Fornasier}, S., {Hasselmann}, P.~H., {Barucci}, M.~A., {et~al.} 2015, \aap,
  583, A30

\bibitem[{{Gr{\"u}n} {et~al.}(2016){Gr{\"u}n}, {Agarwal}, {Altobelli},
  {Altwegg}, {Bentley}, {Biver}, {Della Corte}, {Edberg}, {Feldman}, {Galand},
  {Geiger}, {G{\"o}tz}, {Grieger}, {G{\"u}ttler}, {Henri}, {Hofstadter},
  {Horanyi}, {Jehin}, {Kr{\"u}ger}, {Lee}, {Mannel}, {Morales}, {Mousis},
  {M{\"u}ller}, {Opitom}, {Rotundi}, {Schmied}, {Schmidt}, {Sierks},
  {Snodgrass}, {Soja}, {Sommer}, {Srama}, {Tzou}, {Vincent},
  {Yanamandra-Fisher}, {A'Hearn}, {Erikson}, {Barbieri}, {Barucci}, {Bertaux},
  {Bertini}, {Burch}, {Colangeli}, {Cremonese}, {Da Deppo}, {Davidsson},
  {Debei}, {De Cecco}, {Deller}, {Feaga}, {Ferrari}, {Fornasier}, {Fulle},
  {Gicquel}, {Gillon}, {Green}, {Groussin}, {Guti{\'e}rrez}, {Hofmann},
  {Hviid}, {Ip}, {Ivanovski}, {Jorda}, {Keller}, {Knight}, {Knollenberg},
  {Koschny}, {Kramm}, {K{\"u}hrt}, {K{\"u}ppers}, {Lamy}, {Lara}, {Lazzarin},
  {L{\`o}pez-Moreno}, {Manfroid}, {Epifani}, {Marzari}, {Naletto}, {Oklay},
  {Palumbo}, {Parker}, {Rickman}, {Rodrigo}, {Rodr{\`\i}guez}, {Schindhelm},
  {Shi}, {Sordini}, {Steffl}, {Stern}, {Thomas}, {Tubiana}, {Weaver},
  {Weissman}, {Zakharov}, \& {Taylor}}]{2016Grun}
{Gr{\"u}n}, E., {Agarwal}, J., {Altobelli}, N., {et~al.} 2016, \mnras, 462,
  S220

\bibitem[{{Hansen} {et~al.}(2016){Hansen}, {Altwegg}, {Berthelier}, {Bieler},
  {Biver}, {Bockel{\'e}e-Morvan}, {Calmonte}, {Capaccioni}, {Combi}, {de
  Keyser}, {Fiethe}, {Fougere}, {Fuselier}, {Gasc}, {Gombosi}, {Huang}, {Le
  Roy}, {Lee}, {Nilsson}, {Rubin}, {Shou}, {Snodgrass}, {Tenishev}, {Toth},
  {Tzou}, {Simon Wedlund}, \& {Rosina Team}}]{Hansen2016}
{Hansen}, K.~C., {Altwegg}, K., {Berthelier}, J.~J., {et~al.} 2016, \mnras,
  462, S491

\bibitem[{{Huebner} {et~al.}(1992){Huebner}, {Keady}, \& {Lyon}}]{Huebner1992}
{Huebner}, W.~F., {Keady}, J.~J., \& {Lyon}, S.~P. 1992, \apss, 195, 1

\bibitem[{{Jewitt} \& {Meech}(1986)}]{Jewitt1986}
{Jewitt}, D. \& {Meech}, K.~J. 1986, \apj, 310, 937

\bibitem[{{Kausch} {et~al.}(2015){Kausch}, {Noll}, {Smette}, {Kimeswenger},
  {Barden}, {Szyszka}, {Jones}, {Sana}, {Horst}, \& {Kerber}}]{Kausch2015}
{Kausch}, W., {Noll}, S., {Smette}, A., {et~al.} 2015, \aap, 576, A78

\bibitem[{{Knight} {et~al.}(2017){Knight}, {Snodgrass}, {Vincent}, {Conn},
  {Skiff}, {Schleicher}, \& {Lister}}]{Knight2017}
{Knight}, M.~M., {Snodgrass}, C., {Vincent}, J.-B., {et~al.} 2017, \mnras, 469,
  S661

\bibitem[{{La Forgia} {et~al.}(2019){La Forgia}, {Lazzarin}, {Bodewits},
  {Fulle}, {Bertini}, {Naletto}, {Tubiana}, {Guettler}, {Ivanovsky}, {Sierks},
  \& {Lara}}]{LaForgia2019}
{La Forgia}, F., {Lazzarin}, M., {Bodewits}, D., {et~al.} 2019, in EPSC-DPS
  Joint Meeting 2019, Vol. 2019, EPSC--DPS2019--1442

\bibitem[{{Laeuter} {et~al.}(2020){Laeuter}, {Kramer}, {Rubin}, \&
  {Altwegg}}]{Laeuter2020}
{Laeuter}, M., {Kramer}, T., {Rubin}, M., \& {Altwegg}, K. 2020, arXiv
  e-prints, arXiv:2006.01750

\bibitem[{{McKay} {et~al.}(2015){McKay}, {Cochran}, {DiSanti}, {Villanueva},
  {Russo}, {Vervack}, {Morgenthaler}, {Harris}, \& {Chanover}}]{McKay2015}
{McKay}, A.~J., {Cochran}, A.~L., {DiSanti}, M.~A., {et~al.} 2015, \icarus,
  250, 504

\bibitem[{{Meftah} {et~al.}(2018){Meftah}, {Dam{\'e}}, {Bols{\'e}e},
  {Hauchecorne}, {Pereira}, {Sluse}, {Cessateur}, {Irbah}, {Bureau}, {Weber},
  {Bramstedt}, {Hilbig}, {Thi{\'e}blemont}, {Marchand }, {Lef{\`e}vre},
  {Sarkissian}, \& {Bekki}}]{Meftah2018}
{Meftah}, M., {Dam{\'e}}, L., {Bols{\'e}e}, D., {et~al.} 2018, \aap, 611, A1

\bibitem[{{Morgenthaler} {et~al.}(2001){Morgenthaler}, {Harris}, {Scherb},
  {Anderson}, {Oliversen}, {Doane}, {Combi}, {Marconi}, \&
  {Smyth}}]{Morgenthaler2001}
{Morgenthaler}, J.~P., {Harris}, W.~M., {Scherb}, F., {et~al.} 2001, \apj, 563,
  451

\bibitem[{{Opitom} {et~al.}(2019){Opitom}, {Yang}, {Selman}, \&
  {Reyes}}]{Opitom2019}
{Opitom}, C., {Yang}, B., {Selman}, F., \& {Reyes}, C. 2019, \aap, 628, A128

\bibitem[{{Samarasinha} {et~al.}(2013){Samarasinha}, {Martin}, \&
  {Larson}}]{Samarasinha2013}
{Samarasinha}, N.~H., {Martin}, M.~P., \& {Larson}, S.~M. 2013, {Cometary Coma
  Image Enhancement Facility}, \url{http://www.psi.edu/research/cometimen}

\bibitem[{{Schultz} {et~al.}(1992){Schultz}, {Li}, {Scherb}, \&
  {Roesler}}]{Schultz1992}
{Schultz}, D., {Li}, G.~S.~H., {Scherb}, F., \& {Roesler}, F.~L. 1992, \icarus,
  96, 190

\bibitem[{{Smette} {et~al.}(2015){Smette}, {Sana}, {Noll}, {Horst}, {Kausch},
  {Kimeswenger}, {Barden}, {Szyszka}, {Jones}, {Gallenne}, {Vinther},
  {Ballester}, \& {Taylor}}]{Smette2015}
{Smette}, A., {Sana}, H., {Noll}, S., {et~al.} 2015, \aap, 576, A77

\bibitem[{{Snodgrass} {et~al.}(2017{\natexlab{a}}){Snodgrass}, {Agarwal},
  {Combi}, {Fitzsimmons}, {Guilbert-Lepoutre}, {Hsieh}, {Hui}, {Jehin},
  {Kelley}, {Knight}, {Opitom}, {Orosei}, {de Val-Borro}, \&
  {Yang}}]{Snodgrass2017b}
{Snodgrass}, C., {Agarwal}, J., {Combi}, M., {et~al.} 2017{\natexlab{a}},
  \aapr, 25, 5

\bibitem[{{Snodgrass} {et~al.}(2017{\natexlab{b}}){Snodgrass}, {A'Hearn},
  {Aceituno}, {Afanasiev}, {Bagnulo}, {Bauer}, {Bergond}, {Besse}, {Biver},
  {Bodewits}, {Boehnhardt}, {Bonev}, {Borisov}, {Carry}, {Casanova}, {Cochran},
  {Conn}, {Davidsson}, {Davies}, {de Le{\'o}n}, {de Mooij}, {de Val-Borro},
  {Delacruz}, {DiSanti}, {Drew}, {Duffard}, {Edberg}, {Faggi}, {Feaga},
  {Fitzsimmons}, {Fujiwara}, {Gibb}, {Gillon}, {Green}, {Guijarro},
  {Guilbert-Lepoutre}, {Guti{\'e}rrez}, {Hadamcik}, {Hainaut}, {Haque},
  {Hedrosa}, {Hines}, {Hopp}, {Hoyo}, {Hutsem{\'e}kers}, {Hyland}, {Ivanova},
  {Jehin}, {Jones}, {Keane}, {Kelley}, {Kiselev}, {Kleyna}, {Kluge}, {Knight},
  {Kokotanekova}, {Koschny}, {Kramer}, {L{\'o}pez-Moreno}, {Lacerda}, {Lara},
  {Lasue}, {Lehto}, {Levasseur-Regourd}, {Licandro}, {Lin}, {Lister}, {Lowry},
  {Mainzer}, {Manfroid}, {Marchant}, {McKay}, {McNeill}, {Meech}, {Micheli},
  {Mohammed}, {Mongui{\'o}}, {Moreno}, {Mu{\~n}oz}, {Mumma}, {Nikolov},
  {Opitom}, {Ortiz}, {Paganini}, {Pajuelo}, {Pozuelos}, {Protopapa}, {Pursimo},
  {Rajkumar}, {Ramanjooloo}, {Ramos}, {Ries}, {Riffeser}, {Rosenbush},
  {Rousselot}, {Ryan}, {Santos-Sanz}, {Schleicher}, {Schmidt}, {Schulz}, {Sen},
  {Somero}, {Sota}, {Stinson}, {Sunshine}, {Thompson}, {Tozzi}, {Tubiana},
  {Villanueva}, {Wang}, {Wooden}, {Yagi}, {Yang}, {Zaprudin}, \&
  {Zegmott}}]{Snodgrass2017}
{Snodgrass}, C., {A'Hearn}, M.~F., {Aceituno}, F., {et~al.} 2017{\natexlab{b}},
  Philosophical Transactions of the Royal Society of London Series A, 375,
  20160249

\bibitem[{{Snodgrass} {et~al.}(2016){Snodgrass}, {Jehin}, {Manfroid}, {Opitom},
  {Fitzsimmons}, {Tozzi}, {Faggi}, {Yang}, {Knight}, {Conn}, {Lister},
  {Hainaut}, {Bramich}, {Lowry}, {Rozek}, {Tubiana}, \&
  {Guilbert-Lepoutre}}]{Snodgrass2016}
{Snodgrass}, C., {Jehin}, E., {Manfroid}, J., {et~al.} 2016, \aap, 588, A80

\bibitem[{{Solontoi} {et~al.}(2012){Solontoi}, {Ivezi{\'c}}, {Juri{\'c}},
  {Becker}, {Jones}, {West}, {Kent}, {Lupton}, {Claire}, {Knapp}, {Quinn},
  {Gunn}, \& {Schneider}}]{Solontoi2012}
{Solontoi}, M., {Ivezi{\'c}}, {\v{Z}}., {Juri{\'c}}, M., {et~al.} 2012,
  \icarus, 218, 571

\bibitem[{{Soto} {et~al.}(2016){Soto}, {Lilly}, {Bacon}, {Richard}, \&
  {Conseil}}]{Soto2016}
{Soto}, K.~T., {Lilly}, S.~J., {Bacon}, R., {Richard}, J., \& {Conseil}, S.
  2016, \mnras, 458, 3210

\bibitem[{{Taylor} {et~al.}(2017){Taylor}, {Altobelli}, {Buratti}, \&
  {Choukroun}}]{Taylor2017}
{Taylor}, M.~G.~G.~T., {Altobelli}, N., {Buratti}, B.~J., \& {Choukroun}, M.
  2017, Philosophical Transactions of the Royal Society of London Series A,
  375, 20160262

\bibitem[{{Tubiana} {et~al.}(2011){Tubiana}, {B{\"o}hnhardt}, {Agarwal},
  {Drahus}, {Barrera}, \& {Ortiz}}]{Tubiana2011}
{Tubiana}, C., {B{\"o}hnhardt}, H., {Agarwal}, J., {et~al.} 2011, \aap, 527,
  A113

\bibitem[{{Vincent} {et~al.}(2013){Vincent}, {Lara}, {Tozzi}, {Lin}, \&
  {Sierks}}]{Vincent2013}
{Vincent}, J.~B., {Lara}, L.~M., {Tozzi}, G.~P., {Lin}, Z.~Y., \& {Sierks}, H.
  2013, \aap, 549, A121

\bibitem[{{Weilbacher} {et~al.}(2016){Weilbacher}, {Streicher}, \&
  {Palsa}}]{Weilbacher2016}
{Weilbacher}, P.~M., {Streicher}, O., \& {Palsa}, R. 2016, {MUSE-DRP: MUSE Data
  Reduction Pipeline}, Astrophysics Source Code Library

\end{thebibliography}

\end{document}